\documentclass[twocolumn,tighten]{aastex631}

\usepackage{xcolor}
\usepackage{rotating} 
\usepackage{amsmath}

\graphicspath{{./}{figures/}}

\begin{document}

\title{Dual perspectives on GX 17+2: a simultaneous NICER and NuSTAR study}

\author[0000-0003-0440-7978]{Malu Sudha}
\affiliation{Department of Physics \& Astronomy, Wayne State University, 666 West Hancock Street, Detroit, MI 48201, USA}
\author[0000-0002-8961-939X]{Renee M.\ Ludlam}
\affiliation{Department of Physics \& Astronomy, Wayne State University, 666 West Hancock Street, Detroit, MI 48201, USA}
\author[0000-0001-8371-2713]{Jeroen Homan}
\affiliation{Eureka Scientific, Inc. 2452 Delmer Street, Oakland, CA 94602, USA}


\author[0000-0001-5683-5339]{Dacheng Lin}
\affiliation{Department of Physics, Northeastern University, Boston, MA 02115-5000, USA}
\author[0000-0003-0870-6465]{Benjamin Coughenour}
\affiliation{Space Sciences Laboratory, University of California, 7 Gauss Way, Berkeley, CA 94720-7450, USA}
\affiliation{Department of Physics, Utah Valley University, 800 W. University Parkway, MS 179, Orem, UT 85048, USA}
\author[0000-0002-8294-9281]{Edward M.\ Cackett}
\affiliation{Department of Physics \& Astronomy, Wayne State University, 666 West Hancock Street, Detroit, MI 48201, USA}

\begin{abstract}
We performed the first simultaneous NICER \& NuSTAR spectral and timing study of the Sco-like Z source GX 17+2. The source traced the full Z track during four observations. We detect signatures of relativistic reflection in the broadband spectra and report results using a reflection framework. The disk is relatively close to the innermost stable circular orbit ($\sim$ 1--4 R$_{ISCO}$), which agrees with previous studies of GX 17+2, but the location of the inner disk is farther out in the horizontal branch (HB) and moves inward toward the flaring branch (FB). We find the FB to be the point of closest approach of the disk to the neutron star. We qualitatively conclude that the evolution of the source along the HID is that of a relatively truncated disk in the HB ($\sim$ 4 R$_{ISCO}$) that approaches the neutron star as it goes along the HID towards the normal branch (NB), soft apex (SA), and finally the FB. We attribute the source evolution along the Z track to varying mass accretion rate and disk instabilities. Rms variability increases from the NB towards the SA and then drops to a constant along the FB indicating that the observed variability likely originates from the disk/boundary layer rather than the corona. 

\end{abstract}

\keywords{accretion, accretion disk---binaries: close---stars: individual (GX 17+2)---X-rays: binaries}

\section{Introduction} \label{sec:intro}

In a neutron star (NS) low-mass X-ray binary (LMXB), the NS accretes material from a low mass companion star which leads to the formation of an accretion disk around it that emits in X-rays. These NS LMXBs are classified into Z and atoll sources based on the `Z' or `C' shape they trace out in the hardness intensity diagram (HID) or color-color diagram (CCD). Between the two, Z sources are more X-ray luminous with $L_X~\sim~0.5-1~L_{\rm Edd}$. The Z track traced by these sources have three main branches identified as the horizontal branch (HB), normal branch (NB), and flaring branch (FB) \citep{hasinger1989}. Z sources are further divided into Sco-like and Cyg-like sources based on the shape of the horizontal branch they trace out, with Sco-like sources having a more vertical, short HB and a longer FB compared to Cyg-like sources \citep{kuulkers1995}. The tracks traversed by these sources in their HID/CCD are often attributed to varying mass accretion rates along the different spectral states \citep{hasinger1989,vrtilek1990}. Furthermore, mass accretion rate is also considered to be the parameter that dictates the switch between the two classes of Z and atoll sources, based on studies of XTE J1701-462 \citep{homan2010}.

Several spectral models have been employed in the attempt to model the spectral  behavior of NS LMXBs. There exists a spectral degeneracy, as multiple models predicting distinct accretion disk-corona geometries can fit the spectra equally well. The Eastern model \citep{mitsuda1984} describes a system where the disk emits a multi-temperature blackbody which in turn acts as the supplier of seed photons to a compact corona in the inner disk region that inverse Comptonizes the soft photons from the disk. The Western model \citep{white1986} describes a system that has a single temperature blackbody emission from the NS or close to it and the Comptonized corona in the inner accretion disk region modeled with a powerlaw. \cite{church1995} and \cite{church2004} describe a system with a blackbody emission from the NS surface/boundary layer (BL) and an extended coronal region above the disk that Comptonizes seed photons from the disk. Another model describes the boundary layer as the source of the Comptonized spectrum and the disk as the source of blackbody emission \citep{popham2001}. There are also the hybrid models used by \cite{lin2007} that use a single temperature blackbody and a broken powerlaw for modeling the hard-state spectra and a multitemperature blackbody and single temperature blackbody along with a broken powerlaw for the soft-state spectra. Since all these models can plausibly account for the accretion disk-corona geometry in NS LMXBs, a spectral degeneracy arises, leaving the precise geometry of NS LMXBs an open question.

Furthermore, the accretion disk could be illuminated by a hard X-ray source, which could be the hot corona or the NS surface/boundary layer or a combination of both depending on the state. The reprocessed emission from the accretion disk is most prominently noted as the Fe K$\alpha$ line in the spectrum. This line has an asymmetrically broadened profile and provides us with an estimate of the inner disk radius as this line encompasses the effects of gravitational redshift and Doppler redshift. These effects become more pronounced closer to the NS  \citep{fabian1989}. Therefore, reflection spectral modeling is an effective method for providing an upper limit on the NS radius as well as providing constraints on other spectral parameters (e.g., \citealt{cackett2008}, 
 \citealt{ludlam2017}, \citealt{ludlam2024}).

GX 17+2 is a Sco-like source located at a distance of $\sim$ 13 kpc \citep{galloway2008}. Using Suzaku observations of GX 17+2, \cite{cackett2010} estimated the inclination of the sources to be 15$^\circ$--27$^\circ$ using the \textsc{diskline} model.  Using NuSTAR 3--30 keV data, \cite{ludlam2017} found the inclination to be around 25$^\circ$--38$^\circ$, while \cite{malu2020} found the inclination to be $\sim$ 27$^\circ$ using AstroSAT SXT+LAXPC 0.8--50 keV data. These studies employed reflection modeling using \textsc{reflionx} and \textsc{bbrefl} models respectively. Hence the system has been inferred to be a low inclination system with the inclination angle $<$ 40$^{\circ}$ and with an accretion disk close to the surface of the NS \citep{cackett2010, ludlam2017,agrawal2020,malu2020}. \cite{lin2012} estimated a constant mass accretion rate for GX 17+2 along the Z track. Although the broadband spectrum has been previously explored in the 0.8--50 keV using AstroSAT SXT+LAXPC \citep{malu2020}, a study using the higher energy resolution NICER+NuSTAR data is warranted to better model the reflection spectrum. Here, we study the broadband spectral behavior of the NS LMXB GX 17+2 using simultaneous NICER and NuSTAR observations. We report the NICER+NuSTAR simultaneous reflection spectral studies of GX 17+2.

\section{Observation and Data Reduction}\label{sec:obs}

The simultaneous NICER and NuSTAR observations of GX 17+2 conducted on 2024-03-16, 2024-04-27, 2024-05-27 and 2024-06-08 were used for this study. Observation details are as given in Table \ref{tab1}.
\begin{table*}[!ht]
    \centering
        \caption{Observation Details of GX 17+2.}
    \begin{tabular}{cccccc}
        \toprule
       Obs \# & Mission & Observation ID & Obs. Start Date & Obs. Stop Date & Exp. (ks) \\
        1 & NuSTAR & 30902026002 & 2024-03-16 15:45:07 & 2024-03-17 03:35:05 & $\sim$ 12.6 \\
        &NICER & 7050410101 & 2024-03-16 15:59:39 & 2024-03-16 22:28:05& $\sim$ 4.5 \\
        2 & NuSTAR & 30902026004 & 2024-04-27 09:33:57 & 2024-04-27 21:05:43 &$\sim$ 8.2 \\
        & NICER & 7050410103 & 2024-04-27 08:30:38 &2024-04-27 21:08:20& $\sim$ 6.4 \\
       3 & NuSTAR & 30902026006 & 2024-05-27 06:38:39& 2024-05-27 17:53:31&$\sim$ 8.6 \\
        & NICER & 7050410104 & 2024-05-27 04:02:33 & 2024-05-27 18:02:57 &$\sim$ 12.8 \\
        4 & NuSTAR & 30902026008 & 2024-06-08 06:11:47& 2024-06-08 16:09:08&$\sim$ 9.0 \\
        & NICER & 7050410106 & 2024-06-08 16:01:55 & 2024-06-08 22:30:01 & $\sim$ 1.5  \\
    \end{tabular}

    \label{tab1}
\end{table*}

Standard data reduction procedures were followed for extracting level 1 and level 2 products from both missions using HEASoft 6.34. NuSTAR Data Analysis Software (NuSTARDAS; \citet{nustardas}) was used for extracting NuSTAR level 1 data. NuSTAR pipeline `nupipeline' was used to perform calibration and screening to produce the cleaned event files. The `nuproducts' task was employed to extract the level 2 data products using a 110$^{\prime\prime}$ circular source region and a background region with the same radius far from the source. While extracting products, barycentric correction was performed. Extracted NuSTAR spectra were binned to a minimum of 25 cts per bin.

NICER data products were obtained by performing the `nicerl2' routine. Observations were performed after NICER experienced the light leak\footnote{https://heasarc.gsfc.nasa.gov/docs/nicer/analysis$\_$threads/light-leak-overview/}, therefore we used only the orbit night data. The `barycorr' tool was used for obtaining a barycentered event file. Spectral products were extracted using the `nicerl3-spect' routine and the SCORPEON model was used to estimate the background in our spectral modeling\footnote{https://heasarc.gsfc.nasa.gov/docs/nicer/analysis$\_$threads/scorpeon-overview/}. NICER spectra were optimally binned to a minimum of 10 cts/bin as per the default binning scheme employed in the NICER reduction process. The NICER background lightcurve was estimated using the SCORPEON model. Background counts were estimated to be $<$ 1 count s$^{-1}$, which is negligible with respect to the source count rate and hence were not subtracted (e.g.  \citealt{zhang2021}). All the Fourier timing products such as power density spectra, cospectra and the cross-spectra were obtained using
STINGRAY\footnote{https://docs.stingray.science/en/stable/} \citep{huppenkothen2019,huppenkothen2019a,bachetti2021}. Cross-Correlation functions (CCF) were obtained using the `crosscorr' method in `ftools'.

From now on, we refer to the NuSTAR and NICER observations as obs1, obs2, obs3 and obs4 as shown in Table \ref{tab1}. Lightcurves were analyzed GTI segment-wise for both NICER and NuSTAR data.

\begin{figure*}
\centering
\includegraphics[width=\textwidth, angle=0]{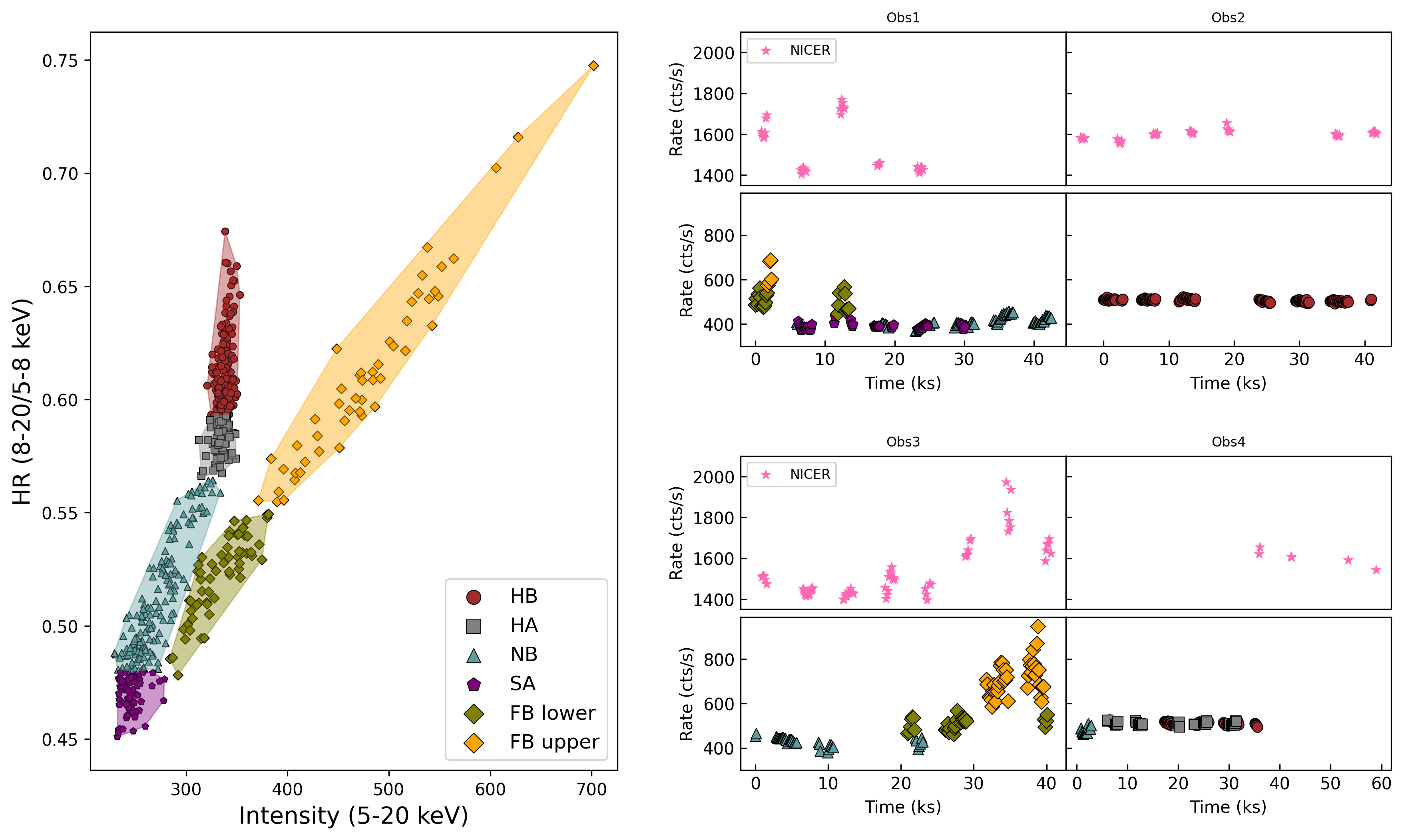}
\caption{Left: HID of GX 17+2 obtained using NuSTAR observations. Hard color is defined as the ratio of X-ray photon counts in the 8--20 keV and 5--8 keV energy bands. Intensity (cts/s) used is from the 5--20 keV energy band. Colored polygons and differently styled markers are used to indicate the  separate branches of the Z track selected for further analysis. Right: NuSTAR lightcurves in the 3--20 keV energy band on the bottom panel (binned to 128 s) using the same styled markers as used in the HID and NICER lightcurves in the 0.5--10 keV energy band binned to 128 s in the top panel marked using the star symbol in pink color.}
\label{fig:hid}
\end{figure*}

GX 17+2 traced out the entire Z track in the HID during these observations. The HID was obtained using NuSTAR lightcurves, with hardness ratio between 5--8 keV and 8--20 keV energy bands and intensity in the 5-20 keV energy band, with a lightcurve bin size of 128 s (see Figure \ref{fig:hid}). The bin size was chosen such that the individual branches could be visually distinguished and separated for further analysis.

Power density spectra (PDS) and co-spectra were generated for each individual GTI based segment of the lightcurve from NICER and NuSTAR respectively. Since NuSTAR observes simultaneously with both FPMA and FPMB instruments, it provides us with an opportunity to produce co-spectra which is the real part of cross spectra. Co-spectra is equivalent to the poisson subtracted PDS and takes deadtime effects into account (see \citealt{bachetti2015} for a detailed explanation). Therefore, we generate co-spectra of the NuSTAR data instead of its PDS. 

Each GTI segment was sectioned into segments of 64s length for which a PDS was generated with a time bin size of 0.0078125 s (i.e., 1/128 s), resulting in a Nyquist frequency and frequency resolution of 64 Hz and 0.015625 Hz (i.e 1/64 s$^{-1}$) respectively. All PDS/cospectra were then combined into a final averaged periodogram/cospectra.  Fractional rms normalization \citep{belloni1990, miyamoto1992} was used to normalize the obtained PDS. NICER background contribution was ignored as it was found to be negligible ($<$ 1 count/s). Poisson noise was subtracted from all the PDS and they were  logarithmically rebinned to lower the noise level. NuSTAR lightcurves used for obtaining the cospectra were background subtracted.

A search for QPOs and QPO-like features was performed in the 0.5--10 keV PDS of NICER observations and the 3-20 keV cospectrum of NuSTAR observations. The PDS and cospectrum were fairly featureless. This is not surprising given the relatively low rms amplitudes of the low- and high-frequency QPOs in Sco-like Z sources \citep{homan2002}. We further determined the integrated fractional rms for each branch of the HID in the 0.01--64 Hz frequency range using the NuSTAR lightcurves.

A cross-correlation function (CCF) study was performed for each GTI segment, by cross-correlating the soft 0.5--3 keV NICER lightcurves with the hard 3--20 keV NuSTAR lightcurves.  Strictly simultaneous GTI segments were created for the CCF analysis by converting  NICER GTIs to NuSTAR times which were then used to extract NuSTAR lightcurves that are strictly simultaneous with NICER lightcurves. A few segments showed positive correlation coefficients indicating a correlated variation between the soft and hard lightcurves, but none of the segments showed any significant soft or hard lags, where soft lags indicate the soft photons lagging behind the hard photons and hard lags vice versa. As per \cite{sriram2019} who attributes the obtained CCF lags to be the readjustment timescales of the Comptonization region, the non-occurrence of lags in our CCF analysis could be indicative of a non-varying coronal region during our observations.

Simultaneous spectra from NICER and NuSTAR observations were extracted to obtain horizontal branch (HB), hard apex (HA), normal branch (NB), soft apex (SA), and flaring branch (FB) spectra for each individual observation. This was achieved by using `niextract-events' to extract the associated NICER events of GTIs corresponding to each branch obtained based on the NuSTAR HID (Figure \ref{fig:hid}). 
\section{Data Analysis \& Results}\label{sec:Results}

XSPEC v12.14.1 was employed to model simultaneous NICER+NuSTAR spectra of all HID branches for each individual observation. NICER spectra was considered in the 1--10 keV energy range and NuSTAR spectra in the 3--25 keV energy range.  C-statistic was employed during fitting, but at higher count rates this approaches $\chi^{2}$ behavior, hence the statistical assumptions remain valid. The spectral continuum was initially modeled following \cite{lin2007} and  \cite{cackett2010}. A multicolor disk blackbody was used for the disk emission, a single temperature blackbody for the boundary layer, a powerlaw for the Comptonized emission, an ISM absorption model (\textsc{Tbabs}), and a multiplicative model to account for the mission specific calibration differences between NICER and NuSTAR called \textsc{CRABCORR} \citep{steiner2010}.  The normalization constant of the \textsc{CRABCORR} model was fixed to 1 for the NuSTAR FPMA spectra and allowed to vary for NuSTAR FPMB and NICER spectra. While the $\Delta \Gamma$ parameter was fixed to zero for FPMA and FPMB spectra, it was fixed to $-0.1$ for the NICER spectra (consistent with the upper bound for other bright XRBs since the light leak in literature; see \citealt{moutard2023, hall2025, ludlam2025}). A more physical model \textsc{thcomp} was also tried to account for the Comptonized emission (see Section \ref{sec:cont} for details). The column density, N$_{H}$, was tied between branches within each observation. NICER spectra below 1 keV were ignored due to large residuals that are instrumental in origin (e.g.  \citealt{li2024}, \citealt{prabhakar2022} (reference therein), \citealt{manca2023}). Since we were not specifically looking for any low energy features, we chose to ignore the $<$ 1 keV energies in the spectra. Furthermore, to verify the reliability of the continuum and disk parameters upon excluding the $<$ 1 keV band, we modeled the obs1 spectral continuum using the best-fit model mentioned in Section \ref{sec:cont} starting from 0.5 keV and determined that the continuum parameters remained within uncertainties, with the fit quality deteriorating with $\Delta$ C-stat $\sim$ 243 for 41 additional degrees of freedom. For NuSTAR spectra, the energy selection was based on where the background dominated. Continuum modeling indicated absorption edges near $\sim$ 1.84 keV and $\sim$ 1.2--1.3 keV and a broad Fe K emission feature at $\sim$ 6.7 keV. The 1.84 keV is assumed to be the Silicon K edge which is a NICER detector feature\footnote{https://heasarc.gsfc.nasa.gov/docs/nicer/analysis\_threads/arf-rmf/} and the residuals around $\sim$ 1.2--1.3 keV could be Mg K edge associated with the ISM \citep{rogantini2020}. To account for the residuals at the absorption edges, two \textsc{edge} models were multiplied to the overall continuum model. 

\begin{figure*}
\centering
  
\includegraphics[width=\textwidth,angle=0]{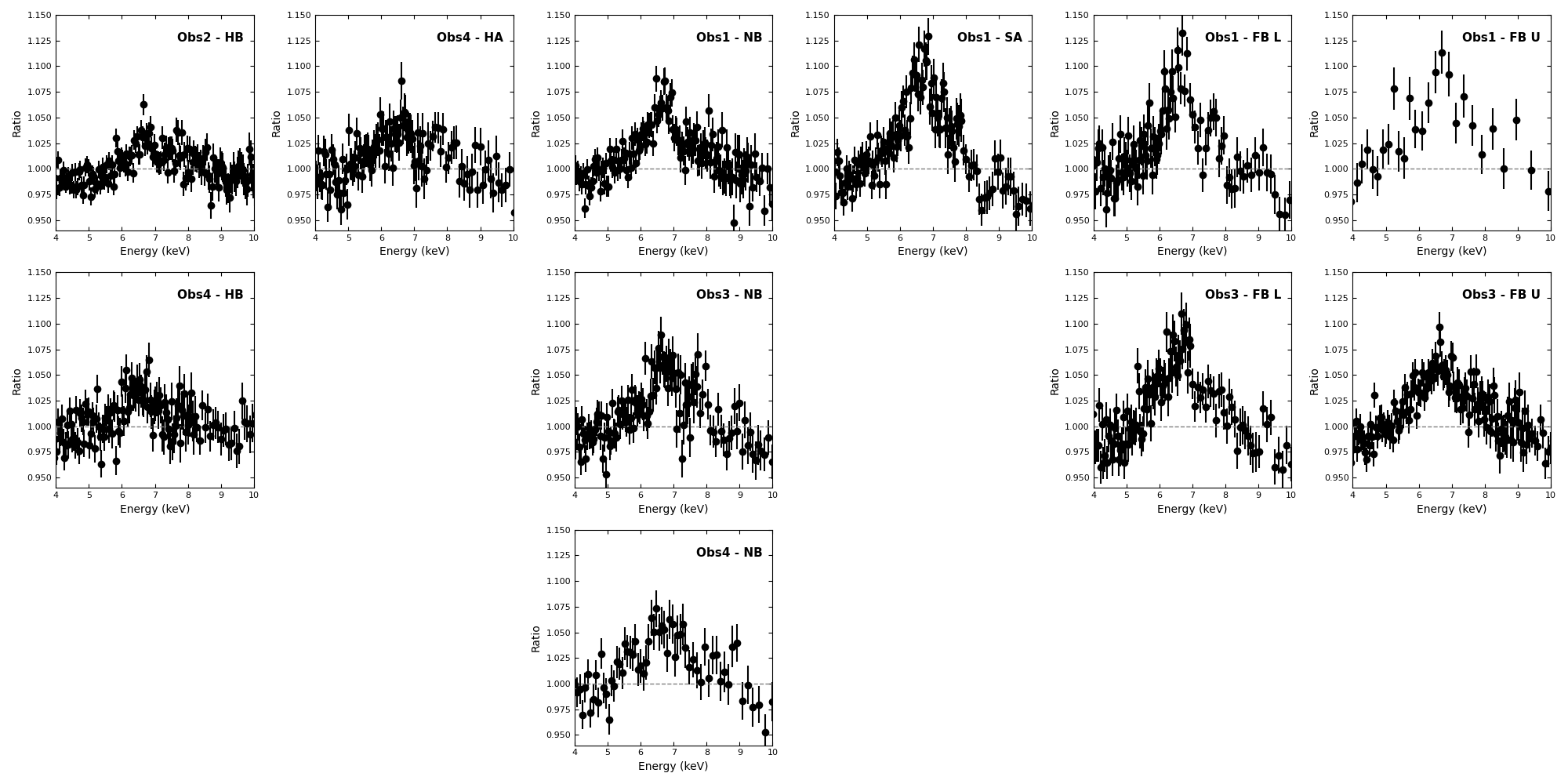}
    \caption{Ratio of NuSTAR FPMA data to continuum model indicating the presence of a Fe K line emission at $\sim$ 6.7 keV. The continuum is modeled by a {\textsc diskbb+bbody+powerlaw} model. Variation in the Fe K strength as it moves along the different branches can be clearly noted.}
    \label{fig:iron}
    \end{figure*}

To account for the signatures of the reflected emission, the \textsc{RELXILLNS} model was used, which is a flavor of the self-consistent reflection model \textsc{RELXILL} \citep{dauser2014,garcia2014}. \textsc{RELXILLNS} considers the source of illumination to be thermal and originating from the boundary layer or the NS surface (\citealt{garcia2022}; for a detailed review and parameter specifics see \citealt{ludlam2024}). A Markov Chain Monte Carlo (MCMC) chain was run with 100 walkers, a chain length of 50000 and burn-in of 10$^{6}$ to estimate errors at the 90\% confidence level. 

Following our data selection, in obs1, the NB, SA, lower FB (FB L), and upper FB (FB U) branches were identified. In obs2, the HB was identified, while obs3 exhibited the NB, FB L, and FB U branches. In obs4, the HB, HA, and NB were identified. All branches were covered by simultaneous NICER and NuSTAR spectra, except the HA and NB in obs4, which were covered by NuSTAR alone. Spectra of all branches in the same observation were fitted simultaneously with parameters such as hydrogen column density, inclination, iron abundance, and emissivity indices tied between branches. We used obs1 as a reference for the spectral modeling of obs2, obs3 and obs4, since obs1 has NB, SA, lower FB and upper FB present in the data, representing the maximum portion of Z track/spectral states covered in a single observation in our entire dataset. The inclination angle and N$_{H}$ were fixed for obs2, obs3 and obs4 based on the value obtained from obs1. The inclination angle obtained from the fit is in agreement with that noted from the literature \citep{ludlam2017,malu2020}. The disk density log (n) was fixed at 19 for all observations, which is the hard limit for disk density in the \textsc{RELXILLNS} model. See Table 2 for the best fit results.
\begin{table*}
\begin{rotatetable*}

\caption{Combined Reflection Modeling for obs1, obs2, obs3 and obs4}
\label{tab:combined-relxillns}
\begin{center}
\scriptsize
\setlength{\tabcolsep}{1pt}
\begin{tabular}{ll|cccc|c|ccc|ccc}
\hline
Model & Parameter & \multicolumn{4}{c|}{Obs1} & \multicolumn{1}{c}{Obs2} & \multicolumn{3}{|c}{Obs3} & \multicolumn{3}{|c}{Obs4} \\ 
&& NB & SA & FB (L) & FB (U) & HB & NB & FB (L) & FB (U) & HB & HA & NB\\
\hline

{\textsc crabcor} 
& $C_{\rm FPMB}$
&$ 0.993 \pm 0.002$ & .. & .. & .. 
&$ 0.982 \pm0.002$ 
&$ 0.984 \pm 0.002$ & .. & .. 
&$ 0.987 \pm0.003$ 
&$ 0.972 \pm0.003$ 
&$ 0.975 \pm0.005$ 
\\

& $C_{\rm NICER}$
&$ 0.760 _{- 0.005 }^{+ 0.004 }$
&$ 0.787 _{- 0.003 }^{+ 0.002 }$
&$ 0.790_{- 0.001 }^{+ 0.005 } $
&$ 0.72 \pm 0.01$
&$ 0.756\pm0.001$ 
&$ 0.741_{- 0.005 }^{+ 0.004}$
&$ 0.799_{- 0.003 }^{+ 0.002}$
&$ 0.698\pm 0.002$ 
&$ 0.769 \pm 0.01 $ \\

& $\Delta \Gamma_{\rm NICER}$ 
& -0.1 (fixed) & .. & .. & .. 
& -0.1 (fixed) 
& -0.1 (fixed) & .. & .. 
& -0.1 (fixed) \\

{\textsc edge}
& E (keV)
&$ 1.24 _{- 0.02 }^{+ 0.02 }$ & .. & .. & ..
&$ 1.27 _{- 0.02 }^{+ 0.00 }$ 
&$ 1.25 _{- 0.02 }^{+ 0.03 } $ & .. & .. 
&$  1.21 _{- 0.04 }^{+ 0.06 }$ \\

& $\tau$ 
&$ 0.08 _{- 0.01 }^{+ 0.01 }$ & .. & .. & ..
&$ 0.09_{- 0.01 }^{+ 0.02} $ 
&$ 0.09 _{- 0.01 }^{+ 0.01 }$ & .. & .. 
&$ 0.23 _{- 0.05 }^{+ 0.09 }$ \\

{\textsc edge}
& E (keV)
&$ 1.84 _{- 0.01 }^{+ 0.01 } $ & .. & .. & ..
&$ 1.85 \pm 0.01 $ 
&$ 1.84 \pm 0.01$ & .. & .. 
&$ 1.84 _{- 0.05 }^{+ 0.02 } $ \\

& $\tau$ 
&$ 0.099 _{- 0.007 }^{+ 0.004 }$ & .. & .. & ..
&$ 0.11 \pm 0.01$ 
&$ 0.11\pm 0.01$ & .. & .. 
&$  0.18 \pm 0.04$  \\

{\textsc \textsc{Tbabs}} 
& $\mathrm{N}_{\mathrm{H}}\ ^{a}$ ($10^{22}$ cm$^{-2}$) 
&$ 3.08 _{- 0.01 }^{+ 0.01 } $ & .. & .. & ..
&$ --$ 
&$ -- $ & .. & ..
&$ --$ \\

{\textsc diskbb} 
& $kT_{\rm in}$ (keV) 
&$ 1.69 _{- 0.03 }^{+ 0.02 }$
&$ 1.78 _{- 0.01 }^{+ 0.01 }$
&$ 2.03 _{- 0.01 }^{+ 0.01 } $
&$ 1.93 _{- 0.05 }^{+ 0.03 }$
&$ 1.74 _{- 0.01 }^{+ 0.05 }$ 
&$ 1.66 _{- 0.02 }^{+ 0.05 }$
&$ 2.07 _{- 0.04 }^{+ 0.03 }$
&$ 2.10 _{- 0.07 }^{+ 0.05 }$ 
&$ 1.68 _{- 0.04 }^{+ 0.03 }$ 
&$1.76 _{- 0.02 }^{+ 0.13 }$
&$1.68 _{- 0.08 }^{+ 0.08 }$\\

& norm$_{\rm disk}$ 
&$113.34 _{- 3.45 }^{+ 6.15 }$
&$ 92.88 _{- 1.71 }^{+ 1.68 }$
&$ 67.72 _{- 1.57 }^{+ 1.06 }$
&$ 86.97 _{- 5.12 }^{+ 6.62 }$
&$  113.18 _{- 9.48 }^{+ 1.33 }$ 
&$ 122.16 _{- 12.25 }^{+ 3.55 } $
&$ 63.62 _{- 3.25 }^{+ 4.11 }$
&$ 69.75 _{- 4.58 }^{+ 6.93 }$ 
&$ 129.54 _{- 9.36 }^{+ 10.94 }$ 
&$ 110.16 _{- 22.33 }^{+ 5.94 }$
&$114.66 _{- 11.38 }^{+ 20.89 }$\\

{\textsc powerlaw} 
& $\Gamma$ 
&$ 2.65 _{- 0.18 }^{+ 0.19 }$
&$ 3.37 _{- 0.16 }^{+ 0.27 }$
&$ 3.65 _{- 0.44 }^{+ 0.34 }$
&$ 2.20 _{- 0.18 }^{+ 0.29 }$
&$2.11 _{- 0.12 }^{+ 0.68 } $ 
&$ 2.82 _{- 0.35 }^{+ 0.63 }$
&$ 3.71 _{- 0.60 }^{+ 0.24 }$
&$ 2.86 _{- 0.66 }^{+ 0.28 }$ 
&$ 2.18 _{- 0.10 }^{+ 0.25 }$ 
&$2.97 _{- 0.09 }^{+ 0.42 }$
&$2.86 _{- 0.17 }^{+ 0.73 }$\\

& norm$_{\rm pl}$ 
&$ 0.45 _{- 0.07 }^{+ 0.02 }$
&$ 0.58 _{- 0.04 }^{+ 0.03 }$
&$ 0.53 _{- 0.09 }^{+ 0.05 } $
&$ 0.18 _{- 0.02 }^{+ 0.02 }$
&$ 0.11 _{- 0.01 }^{+ 0.14 }$ 
&$ 0.38 _{- 0.11 }^{+ 0.17 }$
&$ 0.50 _{- 0.12 }^{+ 0.05 }$
&$ 0.47 _{- 0.11 }^{+ 0.07 }$ 
&$  0.35 _{- 0.06 }^{+ 0.22 }$
&$4.18 _{- 1.02 }^{+ 0.90 }$
&$2.80 _{- 0.75 }^{+ 1.06 }$\\

{\textsc relxillNS} 
& $q$
&$ 2.59 _{- 0.12 }^{+ 0.16 }$ & .. & .. & ..
&$ 2.37 _{- 0.74 }^{+ 7.62 } $ 
&$ 2.73 _{- 0.10 }^{+ 0.24 }$ & .. & .. 
&$ 1.96 _{- 0.59 }^{+ 0.60 } $ 
&$2.57 _{- 0.75 }^{+ 0.57 }$
&$2.57 _{- 0.71 }^{+ 0.94 }$\\

& $i $ ($^{\circ}$)
&$ 37.75 _{- 2.85 }^{+ 0.71 }$ & .. & .. & ..
&$ -- $ 
&$ --  $ & .. & .. 
&$ -- $  \\

& $R_{in}$ (R$_{ISCO}$)
&$ 2.64 _{- 0.35 }^{+ 0.11 } $
&$ 1.74 _{- 0.17 }^{+ 0.18 } $
&$ 1.08 _{- 0.08 }^{+ 0.07 } $
&$ 1.40 _{- 0.04 }^{+ 0.05 }$
&$ 3.83 _{- 2.83 }^{+ 43.36 }$ 
&$ 1.97 _{- 0.76 }^{+ 0.05 } $
&$ 1.21 _{- 0.21 }^{+ 0.75 }$
&$  1.15 _{- 0.15 }^{+ 0.46 }$ 
&$ 1.72 _{- 0.71 }^{+ 0.33 }$
&$2.20 _{- 1.20 }^{+ 1.44 }$
&$1.42 _{- 0.32 }^{+ 0.16 }$\\

& $kT_{bb}$ (keV)
&$ 2.62 _{- 0.04 }^{+ 0.05 }$
&$ 2.80 _{- 0.03 }^{+ 0.05 }$
&$ 3.14 _{- 0.05 }^{+ 0.07 }$
&$ 2.65 _{- 0.08 }^{+ 0.03 }$
&$ 2.73 _{- 0.02 }^{+ 0.08 }$ 
&$ 2.60 _{- 0.02 }^{+ 0.10 } $
&$ 3.16 _{- 0.06 }^{+ 0.06 }$
&$ 2.81 _{- 0.12 }^{+ 0.07 }$ 
&$ 2.65 _{- 0.05 }^{+ 0.09 }$ 
&$2.69 _{- 0.03 }^{+ 0.19 }$
&$2.59 _{- 0.05 }^{+ 0.21 }$\\

& $\log(\xi)$
&$ 2.55 _{- 0.05 }^{+ 0.07 }$
&$ 2.49 _{- 0.02 }^{+ 0.05 }$
&$ 2.56 _{- 0.03 }^{+ 0.06 } $
&$ 2.64 _{- 0.12 }^{+ 0.10 }$
&$ 2.56 _{- 0.05 }^{+ 0.18 }$ 
&$ 2.57 _{- 0.06 }^{+ 0.12 }$
&$ 2.46 _{- 0.08 }^{+ 0.08 }$
&$ 2.67 _{- 0.07 }^{+ 0.04 }$ 
&$ 2.61 _{- 0.06 }^{+ 0.15 }$
&$2.77 _{- 0.43 }^{+ 0.03 }$
&$2.61 _{- 0.04 }^{+ 0.21 }$\\

&$A_{\rm Fe} $
&$ 3.00 _{- 0.25 }^{+ 0.17 }$ & .. & .. & ..
&$ 3.96 _{- 3.00 }^{+ 0.90 }$ 
&$ 3.90 _{- 0.76 }^{+ 0.06 }$ & .. & .. 
&$ 0.60 _{- 0.08 }^{+ 0.11 }$
&$1.01 _{- 0.02 }^{+ 0.02 }$
&$1.13 _{- 0.43 }^{+ 0.15 }$\\

& $\log(n) $ (cm$^{-3}$)
&$ 19 \text{ (fixed)} $ & .. & .. & ..
&$ 19 \text{ (fixed)} $ 
&$ 19 \text{ (fixed)} $ & .. & .. 
&$ 19 \text{ (fixed)} $ \\

& $f_{refl}$
&$ 0.75 _{- 0.06 }^{+ 0.09 }$
&$ 2.67 _{- 0.23 }^{+ 0.20 }$
&$ 10.42 _{- 1.84 }^{+ 1.95 }$
&$  1.02 _{- 0.10 }^{+ 0.07 }$
&$ 0.19 _{- 0.05 }^{+ 0.08 }$ 
&$ 0.54 _{- 0.03 }^{+ 0.18 }$
&$ 6.59 _{- 0.50 }^{+ 2.21 }$
&$0.62 _{- 0.11 }^{+ 0.18 }$ 
&$  0.63 _{- 0.18 }^{+ 0.18 }$ 
&$0.47 _{- 0.09 }^{+ 0.09 }$
&$1.17 _{- 0.43 }^{+ 0.37 }$\\

& norm$_{rel}$ ($10^{-3}$)
&$ 3.81 _{- 0.16 }^{+ 0.23 }$
&$ 1.58 _{- 0.11 }^{+ 0.09 }$
&$ 0.57 _{- 0.10 }^{+ 0.09 }$
&$ 5.81 _{- 0.28 }^{+ 0.53 }$
&$ 8.63 _{- 0.96 }^{+ 0.25 }$ 
&$ 4.80 _{- 0.61 }^{+ 0.16 }$
&$ 0.90 _{- 0.23 }^{+ 0.04 }$
&$ 7.05 _{- 0.99 }^{+ 1.23 }$ 
&$ 6.33 _{- 0.35 }^{+ 0.74 }$ 
&$ 6.76 _{- 1.08 }^{+ 0.29 }$
&$ 4.28 _{- 0.62 }^{+ 0.91 }$\\

\hline
\multicolumn{1}{c}{C-stat (dof)} & \multicolumn{5}{c|}{4734.37 (3967)} & \multicolumn{1}{|c|}{1657.22 (1213)} & \multicolumn{3}{|c|}{4060.19 (3197)} & \multicolumn{3}{|c}{3090.68 (2861)} \\ 
\hline
\end{tabular}

\end{center}

\medskip
\tablecomments{Errors are reported at the 90\% confidence level. The outer disk radius is fixed at 1000 R$_{g}$ and the dimensionless spin parameter is set to $a_{*}=0$ (hence, 1 R$_{ISCO} = 6\ R_{g}$). The notation ``..'' indicates that a parameter is tied across branches for that observation. The notation ``- -'' indicates that the parameter is fixed to the value obtained from obs1.}
\end{rotatetable*}
\end{table*}

\begin{figure*}
\centering
\includegraphics[width=\textwidth, angle=0]{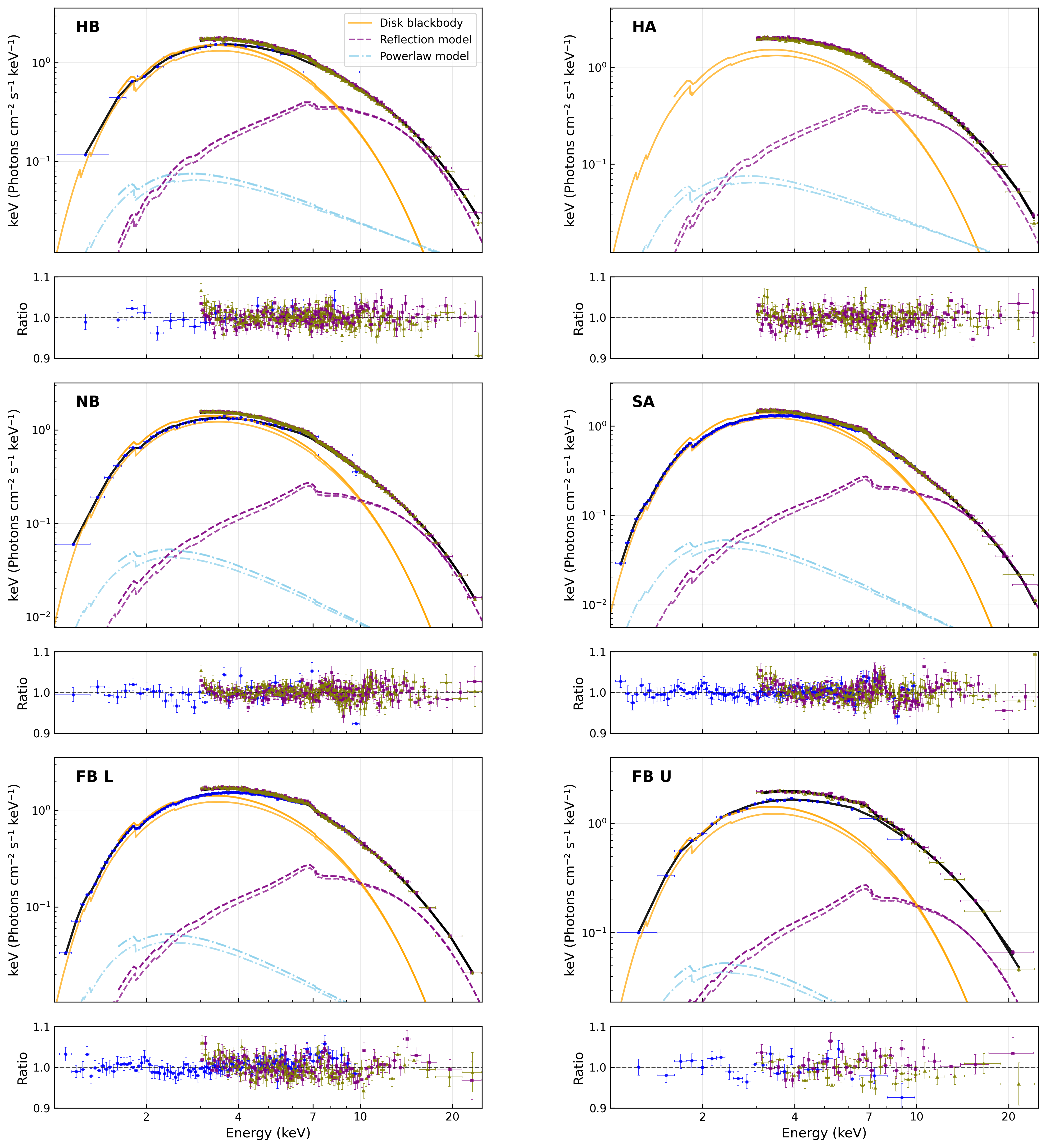}
\caption{NICER + NuSTAR  spectra representative of each branch with the constituent models of the best fit along with the ratio of data to best-fit model. Note that HA is obtained only from NuSTAR observations and it does not have simultaneous NICER observations.}
\label{fig:specmo}
\end{figure*}

\begin{figure}
\centering
  
\includegraphics[width=0.48\textwidth,height=0.8\textwidth]{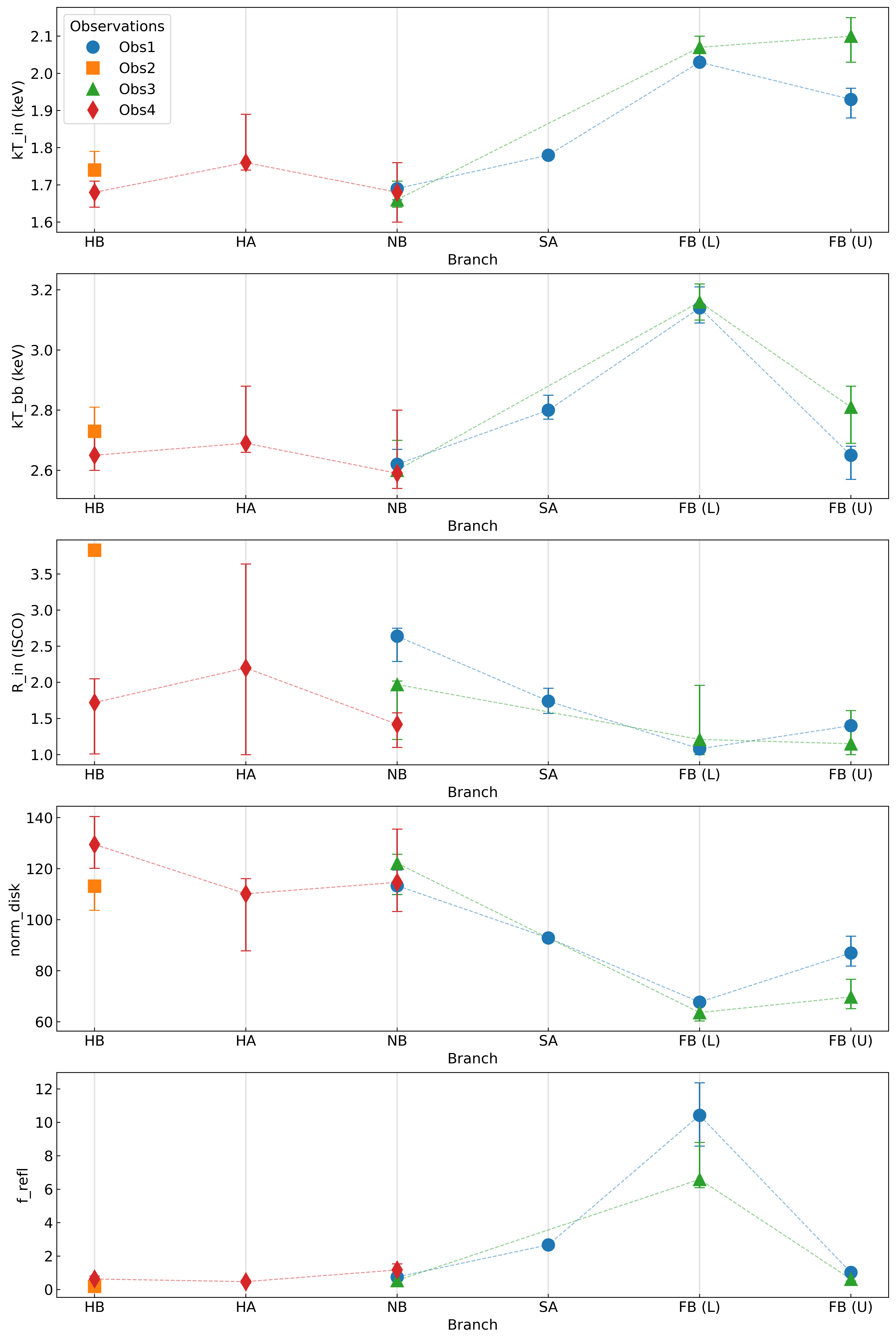}
    \caption{The variation of spectral parameters kT$_{in}$, kT$_{BB}$, R$_{in}$ and \textsc{DiskBB} normalization across the HB, HA, NB, SA, FB lower and FB upper branches. $R_{\rm in}$ for the obs2 HB has been plotted without error bars to aid in proper viewing of all points. This is done so because of the extremely large uncertainty on this point.}
    \label{fig:specpar}
    \end{figure}

\subsection{Continuum modeling}
\label{sec:cont}
As discussed in Section \ref{sec:obs}, the continuum was modeled using an absorbed {\textsc diskbb+bbody+powerlaw} model along with two \textsc{edge} components to account for residuals around 1.2 keV and 1.84 keV. N$_{H}$ obtained for obs1 was used for modeling the remaining observations. The powerlaw index was found to be very soft ($\Gamma$$>$ 4), but it significantly reduced when spectra were modeled with the full reflection framework (discussed in Section \ref{sec:refl}). The continuum model of an absorbed {\textsc diskbb+bbody+powerlaw} yielded a fit statistic of C-stat/degrees of freedom 7246.40/3982. We also tried modeling the continuum with a cut-off powerlaw model instead of a simple powerlaw, but it did not result in meaningful constraints with the model being insensitive to the high energy cut-off parameter value, tending to $<$ 1 keV.

Additionally, we tried the more physical convolution model \textsc{thcomp} \citep{zdziarski2020} to account for the Comptonized emission. For model computation the energy range was extended from 0.1 keV to 1000 keV with 1000 logarithmic bins as recommended when using the model. This convolution model was applied separately on both the disk blackbody and single temperature blackbody components. While using the \textsc{thcomp} model, the optical depth parameter was used instead of the photon index parameter and it was fixed at a value of $\tau$ = 10 (e.g. \citealt{chattopadhyay2024}). We also tested fits where the optical depth parameter was allowed to vary, and alternatively tested fits using the  photon index parameter instead of the optical depth parameter. The remaining parameters of the model, i.e. covering fraction and electron temperature, were free to vary. When the thermal Comptonization model was involved in the fit with the disk blackbody model being convolved, it resulted in a much higher fit statistic of 10682.28/3982, whereas that with the convolution model applied on the single temperature blackbody component gave a lower fit statistic of 8120.65/3982. But as noted from the fit statistic, the addition of the thermal Comptonization models gave worse fits than with the simple powerlaw model. 

The statistical metric of the Bayesian Information Criterion (BIC) was estimated for each continuum fit with  BIC = 10961.73, 8402.83, and 7528.53 for the continuum models involving \textsc{thcomp} convolved with \textsc{DiskBB}, \textsc{thcomp} convolved with \textsc{bbody}, and the continuum model with the simple powerlaw, respectively. The model with the simple powerlaw index is statistically favored over the two Comptonized models. To strengthen the case, we complement the BIC metric with the Akaike information criterion (AIC) metric for the different continuum models. The AIC metric also favors the continuum model with the simple powerlaw having AIC $\sim$ 7314.40, while the \textsc{thcomp} convolved with \textsc{DiskBB} and \textsc{thcomp} convolved with \textsc{bbody} had AIC estimates of 10747.60 and 8188.70, respectively. Furthermore, it was noted that for both the thermal Comptonization models, the covering fraction tended to 1 and the electron temperature was very high ($\sim$ 100--140  keV) indicating unphysical constraints on these parameters. As indicated by the initial powerlaw fit in the continuum model, it is evident that the powerlaw slope is very high, thus mimicking a soft component. This is possibly why the electron temperature is not well constrained by the \textsc{thcomp} model. This is likely caused by the lack of data above 25 keV, which is typically where the Comptonizing component dominates.

Since \textsc{thcomp} did not yield meaningful constraints on the parameters, we continue with the simple powerlaw model to model the Comptonized emission for further modeling. The plot of ratio of data to continuum model focusing on the Fe K region is shown in Figure \ref{fig:iron}. It can be clearly noted from the ratio plots that the iron line strength varies along the branches. HB shows the weakest Fe K line, while the feature consistently increases in strength towards HA, NB, SA and FB (discussed further in Section \ref{sec:disc}).
 
\subsection{Reflection modeling}
\label{sec:refl}
As discussed in Section \ref{sec:obs}, \textsc{RELXILLNS} was used for modeling the reflection spectrum, which considers the thermal emission from NS/BL to be illuminating the accretion disk. Our spectral analysis using the self-consistent \textsc{RELXILLNS} model indicates an inclination of 37$^\circ$$_{- 3 }^{+ 1 }$, which is in close agreement with the previously reported values \citep{cackett2010, ludlam2017,malu2020}. 

A single emissivity profile (i.e. $q1=q2$) is assumed as is typically the case for NS LMXBs. The parameter range of the emissivity index is within the usual range of 1.5--4 seen in NS LMXBs \citep{ludlam2024}. The disk density was fixed at 10$^{19}$ cm$^{-3}$ as previously mentioned in Section \ref{sec:obs}. This is done since the disk density is expected to be $>$ 10$^{20}$ cm$^{-3}$ for the disk around a NS \citep{shakura1973, frank2002}. Hence we resort to using the model upper limit of 10$^{19}$ cm$^{-3}$.

The inner disk radii (R$_{in}$) estimates indicate that the disk is relatively close to the R$_{ISCO}$, with a higher R$_{in}$ in the HB to the lowest in the FB (Figure \ref{fig:specpar}). The reflection fraction is found to be very high in FB (L) ($\sim$ 6--10). Very high values of the reflection fraction is possible in cases when the source that is illuminating the accretion disk is very close to the disk and is located deeper in the gravitational potential well, such that light bending effects are high owing to the majority of the emitted photons interacting with the disk and very few being directly emitted \citep{dauser2016}.
As shown in Figure \ref{fig:specpar}, we also note that the disk  temperature in HB and upper FB seem to be slightly different between observations with HB having a disk temperature of $kT=1.74 _{- 0.01 }^{+ 0.05 }$ keV in obs2 and $kT=1.68 _{- 0.04 }^{+ 0.03 }$ keV in obs4 and FB having a disk temperature of $kT=1.93 _{- 0.05 }^{+ 0.03 }$ keV in obs1 and $kT=2.10 _{- 0.07 }^{+ 0.05 }$ keV in obs3. For NB it is consistent within error bars, with disk temperatures of $kT=1.69_{-0.03}^{+0.02}$ keV in obs1, $kT=1.66_{-0.02}^{+0.05}$ keV in obs3, and $kT=1.68\pm0.08$ keV in obs4. This seems to be the same case for the blackbody temperature and the disk normalization parameters. We note a spread in the inner disk radii values in the NB, which could be due to the observations sampling slightly different parts of the NB, along with the consideration that constraints on NB in obs4 comes only from the NuSTAR spectrum. Reflection fraction on the other hand seems to be fairly consistent for each branch between observations. These results do not conclusively indicate any significant temporal variation in parameter values. Results obtained from the best fit model incorporating the \textsc{RELXILLNS} model are given in Table \ref{tab:combined-relxillns}. See Figure \ref{fig:specmo} for the best fit model and ratio of data to model.

\section{Discussion} \label{sec:disc}

\subsection{Iron line variation along branches: Interpreting the physical picture}

We performed the first ever simultaneous NICER+NuSTAR broadband spectral analysis of GX 17+2 using a reflection framework. We clearly note a variation in the Fe K line strength across branches, being the lowest in the HB to peaking in the FB (see Figure \ref{fig:iron}). The reflection effects would be more pronounced closer to the NS. This is further supported by the trend in spectral parameters including R$_{in}$ and reflection fraction. R$_{in}$ appears to reach its lowest value during the FB, suggesting that the disk extends closest to the NS in this state. However, the changes in R$_{in}$ observed along the Z track are not statistically robust, and thus, the results should be interpreted with caution. The qualitative picture that emerges from the spectral parameter trends is the following:\newline

1. A relatively truncated accretion disk is noted in the HB as indicated by the higher R$_{in}$. However, the constraints on the inner disk radius are weakest in the HB, and thus this interpretation should be treated with caution. Disk and blackbody component shows relatively lower temperature values which is consistent with the disk being relatively further away from the NS. Furthermore, HB has a lower reflection fraction that supports the above picture.   \newline

2. A gradual increase in disk and blackbody temperatures is noted, combined with a decreasing R$_{in}$ and higher reflection fraction going down the HID, towards NB, SA, and FB. Considering obs1, which has the source traversing from NB to FB, we note this evolution in the context of total mass accretion rate $\dot{m}$ as well. In the NB and SA we have a comparatively lower  $\dot{m}$ of 2.17--2.14$\times$10$^{18}$ g s$^{-1}$ (estimated from 1--25 keV unabsorbed flux from spectra), whereas in the lower FB we see an increased $\dot{m}$ of 2.61$\times$10$^{18}$ g s$^{-1}$. This indicates that the mass accretion rate slightly increases down the HID (from NB to FB), leading to the disk further approaching the NS. We refrain from comparing mass accretion rates between branches of different observations as the observations are each a month apart, thus preventing a direct comparison. We also note that extending the energy range for flux estimation does not lead to any significant difference in the flux values, especially since GX 17+2 is a spectrally soft source and the contribution from the hard energies is weak.\newline

3. The lower FB has the maximum disk and blackbody temperatures along with the highest reflection fraction combined with the lowest value of R$_{in}$. We could consider that this is the location that marks the onset of flaring followed by unstable nuclear burning \citep{church2012} and the point of closest approach of the disk to the NS. \newline

4. The upper FB shows a slightly higher R$_{in}$ and much small reflection fraction compared to the lower FB despite the higher $\dot{m}$ (=3.21$\times$10$^{18}$ g s$^{-1}$), possibly owing to the disk receding due to dissipation of energy via flaring and also radiation pressure from the central source. Slight decrease in temperatures could be an indication of fast radiative cooling.

\subsection{Inner disk radius constraints}
As per \cite{mitsuda1984}, \textsc{DiskBB} normalization is given by N$_{dBB}$ = 
(R$_{in}$/D$_{10}$)$^{2}$ $\times$ cos $i$. Here D$_{10}$ stands for the distance to the source (units of 10 kpc) and  R$_{in}$ the apparent inner disk radius. Based on spectral modeling we consider an inclination angle (i) of $\sim$ 37.75$^{\circ}$ and estimate the apparent R$_{in}$ to be 15.18--15.99 km, 13.79--14.23 km, 11.76--12.14 km and 13.15–14.16 km in obs1 NB, SA, FB L and FB U respectively when propagating errors. We must apply the correction factors for spectral hardening $\kappa$ and inner boundary condition $\xi$ to the apparent  R$_{in}$ to obtain the effective R$_{in}$.  As per \cite{shimura1995} and \cite{kubota1998},  $\kappa$ value is $\sim$ 1.7 and $\xi$ value is 1, respectively. We consider the effective R$_{in}$ to be $\kappa^{2}$ $\xi$ R$_{in}$ \citep{kubota2001}. Therefore, the effective R$_{in}$ for  obs1 NB is 44--46 km, SA is 40--41 km, FB L is 34--36 km, and FB U is 38--41 km. Similarly for HB in obs2, effective R$_{in}$ is estimated to be 43--45 km. NB for obs2 has effective R$_{in}$ $\sim$ 44--47, while FB L and FB U has effective R$_{in}$ $\sim$ 33--35 km and 34--37 km respectively. Obs4 HB has effective R$_{in}$ $\sim$ 46--50 km, with HA and NB having effective R$_{in}$ $\sim$ 40--46 km and 43--49 km respectively. 

The radius of the innermost stable circular orbit (ISCO) around the NS in units of gravitational radii R$_{g}$ has a dependency on the 
dimensionless spin parameter $a_{*}$, which is assumed to be 0. With spin parameter $a_{*}=0$, we get 1 R$_{ISCO} = 6\ R_{g}$ where R$_{g}$ is GM/c$^{2}$.
This helps us convert the inner disk radius obtained from \textsc{RELXILLNS} in units of ISCO to that in gravitational radii units. Then, using the NS mass estimate of 1.4 M$_\odot$, we convert R$_{in}$ from units of gravitational radii to km.
Therefore, for obs1 we obtain R$_{in}$ of $32.8_{-4.3}^{+1.4}$ km for NB, $21.6_{-2.1}^{+2.2}$ km for SA, $13.4_{-1.0}^{+0.9}$ km for FB L and $17.4_{-0.5}^{+0.6}$ km for FB U. For obs2 we obtain R$_{in}$ of $47.5_{-35.1}^{+537.9}$ km for HB, while for obs3 we obtain R$_{in}$ of $24.4_{-9.4}^{+0.6}$ km for NB, $15.0_{-2.6}^{+9.3}$ km for FB L and 
$14.3_{-1.9}^{+5.7}$ km for FB U. For Obs4 we obtain R$_{in}$ of $21.3_{-8.8}^{+4.1}$ km for HB, $27.3_{-14.9}^{+17.9}$ km for HA and $17.6_{-4.0}^{+2.0}$ km for NB. Therefore, the values of R$_{in}$ obtained from \textsc{DiskBB} are in close agreement with R$_{in}$ obtained from the \textsc{RELXILLNS} model, taking into account that the R$_{in}$ derived from the \textsc{DiskBB} model can exceed the true inner disk radius by a factor of $\sim$2.2, owing to the assumptions of a geometrically thin disk and zero-torque inner boundary conditions \citep{zimmerman2005}.
Therefore, the inner disk radius values obtained from the \textsc{DiskBB} model are in agreement with the scenario of the disk being close to the R$_{ISCO}$ as suggested by our reflection spectral modeling as well as the literature. Furthermore, the agreement between the radius measurements from the different spectral components establishes the validity of the spectral modeling results.

Given that the disk is truncated, we estimate the maximum radial extent of a BL extending from the NS surface using the obtained mass accretion rate as per \cite{popham2001}. 
We used the equation,
\begin{equation}
\log(R_{BL} - R_{NS}) \sim 5.02 + 0.245 \Bigg[\log\Big({\frac{\dot{M}} {10^{-9.85}\ \ M_{\odot} yr^{-1}}\Big)}\Bigg]^{2.19}
\end{equation}
to estimate the BL radius for all branches in obs1 to understand and compare with other radius constraints. We estimate the R$_{BL}$ to be $\sim$ 59 km, 57 km, 75 km, and 97 km for NB, SA, FB L, and FB U, respectively. The BL radii thus estimated are larger than the inner disk radii estimated from reflection modeling, as well as the \textsc{DiskBB} model.

The inner disk may be truncated by the effects of the magnetic field of the NS. Hence by using the obtained range of R$_{in}$, we estimate the magnetic field strength of the system. We use the equation from \cite{cackett2009},

\begin{equation}
\begin{split}
\mu =\; & 3.5 \times 10^{23} \, k_A^{-7/4} \, x^{7/4} 
\left( \frac{M}{1.4 \, M_\odot} \right)^2 \\
& \times \left( \frac{f_{\mathrm{ang}}}{\eta} \cdot \frac{F_{\mathrm{bol}}}{10^{-9} \, \mathrm{erg} \, \mathrm{cm}^{-2} \, \mathrm{s}^{-1}} \right)^{1/2}
\frac{D}{3.5 \, \mathrm{kpc}} \, \mathrm{G \, cm}^3
\end{split}
\end{equation}
where R$_{in}$=$x$GMc$^{-2}$ to estimate magnetic dipole moment $\mu$. We assume $\eta$ = 0.2 \citep{sibgatullin2000}, conversion factor (k$_A$) =1, and angular anisotropy f$_{ang}$ = 1 as in \citet{ludlam2019}. By taking the range from the highest to lowest R$_{in}$ from HB to FB low and the corresponding bolometric flux in 0.01--100 keV energy range, we estimate the upper limit of 
$\mu$ to be 3.71 $\times$ 10$^{27}$ G cm$^3$ respectively. A distance of 13 kpc was considered for the system. Assuming a radius of 10 km for the NS, the magnetic field strength would be $\sim$ 0.76--7.42  $\times$ 10$^{9}$ G at the magnetic poles of the NS. However, there are no pulsations seen to indicate magnetospheric truncation.
We must note here that this estimate of the upper limit of magnetic field strength is dependent on many assumptions. Primarily, this estimate stems from the assumption that the accretion disk is truncated by the magnetic field at the magnetospheric radius, which might not be the case here. Furthermore, the correction factor k$_A$ can vary between 0.5--1.1 (\citealt{long2005}, \citealt{ibragimov2009} and references therein) and as noted by \citet{cackett2009}, k$_A$ = 0.5 can increase the magnetic field strength estimate by a factor of $\sim$ 3.

\subsection{Evolution along the Z track}

\citet{hasinger1990} used multiwavelength observations of Cyg X-2 to suggest that the mass accretion rate monotonically increases along the Z track from HB to FB. This finding was challenged by observations indicating a decrease in intensity as Z sources moved from NB towards FB. Later, with the advent of RXTE and BeppoSAX offering higher quality data, it enabled a more refined approach of understanding spectral evolution in these sources. Based on the 0.1-200 keV flux on the NB and HB of GX 17+2, \citet{homan2002} suggested that mass accretion rate increases in the opposite direction or that it does not change at all along the Z track owing to only moderate changes in the flux. This led to the conclusion that motion along the Z track could be due to another parameter such as the inner disk radius. While generally the NS LMXB spectra was modeled using a two-component Eastern or Western model (see Section \ref{sec:intro}), \citet{church2006} proposed a Comptonization component from an extended accretion disk corona instead of a compact central corona. Using this model on GX 340+0, they found the mass accretion rate to be increasing in the opposite direction with the minimum being in the FB and the mechanism in FB was proposed to be unstable nuclear burning. Using the hybrid model proposed by \citet{lin2007}, where the spectra is modeled using the two thermal models and a Comptonization model (cutoff powerlaw), \citet{lin2012} modeled GX 17+2 to find that mass accretion rate remained constant along the Sco-like Z track.


More recently, \cite{bhattacherjee2025} performed a broadband spectral study using AstroSAT (NB and FB) and NICER data (FB). They found that the decrease in fraction of disk photons entering the corona, f$_{s}$, was the main cause for the source moving along the NB, while f$_{s}$ remained constant in the FB. Mass accretion rate and T$_{in}$ showed an increase from SA to FB, while they remained constant in NB. Using their primary model (absorbed single temperature blackbody and disk blackbody convolved with \textsc{thcomp}), they found a decrease in R$_{in}$ along NB, while R$_{in}$ remained constant in FB. Their alternate model (absorbed disk blackbody and single temperature blackbody convolved with \textsc{thcomp}) indicated a constant R$_{in}$ throughout the track.

To discuss the evolution of the source along the Z track, there is a need to trace the evolution of the different spectral components along the track. This is dependent on the choice of the spectral model adopted for this purpose. We adopt a full reflection framework for our analysis. The full reflection model self-consistently accounts for both the reprocessed continuum and the superimposed emission lines and absorption edges.
Based on our spectral analysis, as mentioned in the previous section, we note an increase in $\dot{m}$ going from NB to FB (obs1). Although we observe a slight variation in R$_{in}$, it is not statistically significant enough to attribute motion along the Z track to changes in R$_{in}$. Apart from the total mass accretion rate discussed in the above section, the $\dot{m}$ associated with the disk component was inferred from the standard disk equation  (e.g., \citealt{lin2009} and references therein). We notice an almost consistent $\dot{m}$ from NB to SA, and an increase in $\dot{m}$ from the lower FB to the upper FB. This increase in mass inflow rate could suggest an increase in the radiation pressure on the inner disk, thus supporting the idea of a radiation pressure–driven disk in the FB causing disk instabilities. Although this scenario is plausible, it is also likely that the modest increase in $\dot{m}$ alone cannot fully account for the FB evolution.

Based on our temporal analysis along the Z track, we obtained integrated fractional rms in the frequency range of 0.01--64 Hz for each branch of the HID using the NuSTAR lightcurves. We estimated an integrated fractional rms of $\sim$ 1.09\% in the NB, 1.90\% in the SA, 1.27\% in the lower FB, and 1.32\% in the upper FB. Although very low in strength, rms variability increases from the NB towards the SA and then drops to a near constant along the FB. This behavior, combined with the physical picture of increasing soft component contribution down the HID, suggests a straightforward interpretation that the observed variability likely originates from the disk or boundary layer rather than the Comptonized corona. This is further supported by the lack of CCF lags that are attributed to a varying corona \citep{sriram2019}. 

Qualitatively we could conclude that the source evolves along the Z track due to a varying mass accretion rate and disk instabilities. 

Although this is qualitatively compatible, the lack of significantly strong rms variability and CCF lags, coupled with the findings of only marginal changes in R$_{in}$ indicate that evolution along the FB might not be uniquely driven by radiation pressure. A combination of local instabilities or geometrical changes in disk-corona coupling could also be a possibility.

\section{Conclusion}
We performed the first simultaneous NICER+NuSTAR spectral and timing study of the Z source GX 17+2 using a full reflection framework. During the four sets of observations, the source traced out the complete Z track and traversed from HB to FB. We summarize the results as follows: \\

1. Spectral modeling using the full reflection framework revealed that the disk is relatively close to the R$_{ISCO}$, which is in close agreement with previous studies, but with the highest disk truncation noticed in the HB. 

2. The above scenario agrees with our findings that the iron line strength varies along the HID, with the lowest strength being in the HB and the highest in the FB.

3. Considering obs1 that has the source traversing from NB to lower FB, a gradual increase in disk and blackbody temperatures are noticed with a decreasing R$_{in}$ and higher reflection fraction going down the HID.

4. Based on the parameter values in the lower FB, we notice that it has the highest reflection fraction, lowest value of R$_{in}$, and highest values of disk and blackbody temperatures. We qualitatively conclude that this location in the HID marks the onset of flaring followed by unstable nuclear burning and the point of closest approach of the disk to the NS.

5. We propose that the qualitatively concluded evolution of the source along the spectral states is that of a relatively truncated disk in the HB that approaches the NS as it goes along the HID towards the NB, SA, and finally the FB, where the flaring begins. From here the disk slightly recedes as the energy is dissipated and due to effects of radiation pressure. Furthermore, CCF analysis indicates a non-varying coronal region.

6. Spectral and temporal studies indicate that the evolution of the source along the Z track could possibly be attributed to varying mass accretion rates and disk instabilities. But the possibility of a combination of local instabilities or geometrical changes in disk-corona coupling cannot be ruled out.

Similar broad-band spectral reflection spectral modeling are crucial to estimate disk radii constraints and get a better overall understanding of the accretion disc geometry in NS LMXBs.\newline

\section{Acknowledgements}
We thank the anonymous referee for their detailed feedback that has enhanced the scientific rigor of the manuscript. This research has made use of data and/or software provided by the High Energy Astrophysics Science Archive Research Center (HEASARC), which is a service of the Astrophysics Science Division at NASA/GSFC. This research has made use of data from the NuSTAR mission, a project led by the California Institute of Technology, managed by the Jet Propulsion Laboratory, and funded by the National Aeronautics and Space Administration. Data analysis was performed using the NuSTAR Data Analysis Software (NuSTARDAS), jointly developed by the ASI Science Data Center (SSDC, Italy) and the California Institute of Technology (USA). This work was supported by Caltech and JPL through award 1713290.

\end{document}